\documentclass{ws-procs975x65}
\usepackage{graphicx}

\begin{document}

\title{CASIMIR INTERACTION\\ BETWEEN ABSORBING AND META MATERIALS}

\author{FRANCESCO INTRAVAIA\footnote{Supported by QUDEDIS (ESF program) and FASTNet (European Research Training Network).}  and CARSTEN HENKEL}

\address{Institut f\"ur Physik, Universit\"{a}t Potsdam, 
    14469 Potsdam, Germany\\
\email{francesco.intravaia@physik.uni-potsdam.de}}

\begin{abstract}
We investigate the Casimir energy between two dissipative mirrors in
term of a sum over mode formula which can be interpreted by analogy to
a quantum dissipative oscillator.  We also show that metamaterials
engineered at scales between the nanometer and the micron seem a
promising way to achieve a repulsive force.
\smallskip

\end{abstract}
\bigskip

\noindent The Casimir force is one of the most accessible experimental consequences 
of vacuum fluctuations in the macroscopic world. It is the most 
significant force between neutral, non-magnetic objects at 
distances on the micrometer scale and below. For many
experiments searching for novel short-range forces predicted by
unification models \cite{Onofrio:2006, Lamoreaux:2005, 
most}, theoretical calculations of the Casimir force are 
crucial and have to be done at the same level of precision as the
experiments \cite{astrid:epjd}. 

In this context, it is essential to account for the differences 
between the ideal Casimir case and real-world experiments, for
example non-perfect reflectors made from absorbing material.
This problem, in the plate-plate geometry, was solved by 
Lifshitz\cite{lifshitz}
\begin{equation}
    F(L) = 2 \hbar \,{\rm Im} \int\limits_{0}^{\infty}
    \frac{ {\rm d}\omega }{ 2\pi }
    \coth( {\textstyle\frac12} \beta \omega )
    \int\limits_{0}^{\infty}
	\frac{ {\rm d}k\, k k_{z}}{ 2\pi }
	\sum_{\lambda = {\rm TE,\,TM}}
	\left( \frac{ {\rm e}^{ 2 {\rm i} k_{z} L } }{
	r_{\lambda,1} r_{\lambda,2} } - 1
	\right)^{-1}, 
    \label{eq:Lifshitz}
\end{equation}
where $L$ is the distance between the plates,
$\beta = \hbar / k_{B} T$, $k_{z} = (\omega^2/c^2 - k^2)^{1/2}$,
and $r_{{\rm TE},\alpha}$, $r_{{\rm TM},\alpha}$ are the reflection
coefficients at plate $\alpha = 1,2$ for the two principal
polarizations of the electromagnetic field. 
This formula allows to calculate the Casimir force (per unit area)
in terms of the optical properties of the plates, with any non-ideal
behaviour (finite permittivity, dissipation) taken into 
account by suitable models for the reflection coefficients. 
For example, with a dissipative medium one uses a complex dielectric
function provided it is compatible with causality constraints
\cite{landaulif,jackson}.

Lifshitz' approach rather differs from the one used by Casimir. In 
fact, Casimir summed the zero-point energies of the electromagnetic 
modes inside a cavity of perfectly reflecting mirrors (Dirichlet 
boundary conditions), renormalizing this sum by removing the
free vacuum energy. The Casimir energy for a cavity with real mirrors
can also be obtained in this way, the modes being here the ones of
the real cavity\cite{schram}. Adopting a dissipation-less model for the
dielectric function, the final expression coincides with Lifshitz'
formula~(\ref{eq:Lifshitz}). Lifshitz theory has, however, a wider range
of applicability because dissipative mirrors can also be described.
We have shown that, also in this case, the 
Casimir effect can be expressed as a sum over modes.\cite{FI-CH-prep}
A calculation along lines similar to 
Ref.\citen{Intravaia05} allows to transform Eq.\eqref{eq:Lifshitz} 
into (zero temperature, identical mirrors)
\begin{equation}
F(L) = \frac{ \partial }{ \partial L }
\frac{ \hbar }{ 2Ê}
\int\limits_{0}^{\infty}
    \frac{ {\rm d}k\, k }{ 2\pi }
   {\rm Re}\left[ \sum_{n\lambda}
    \omega_{n\lambda}( k ) 
    -2 {\rm i} \frac{\omega_{n\lambda}( k )}{\pi}
    \log\frac{ \omega_{n\lambda}( k ) }{ \omega_{c} }
    \right]^{L}_{L\rightarrow \infty},
    \label{eq:sum-modes}
\end{equation}
where $\omega_{c}$ is an arbitrary cutoff frequency and the discrete 
index $n$ labels the different modes that exist for a given 
$k$-vector 
and polarization. The frequencies $ \omega_{n\lambda}( k )$ are 
the complex solutions of
$ {\rm e}^{ 2 {\rm i} k_{z} L }/ r_{\lambda}^2 - 1=0$.
Their imaginary parts obey a specific sum rule that removes the
dependence on the cutoff $\omega_{c}$.
The result~(\ref{eq:sum-modes}) can be understood by analogy to the
quantized oscillator coupled to a bath, establishing a bridge between 
the quantum field theory and the theory of open systems.  At zero 
temperature, the oscillator's zero-point energy is shifted because the 
bath quantum fluctuations couple to the oscillator observables 
\cite{Nagaev02}. The logarithmic term arises because the ground state 
of the uncoupled oscillator is no longer an eigenstate for the whole 
system, therefore its energy shows fluctuations.

In the non-dissipative case and at short distance, it is well known 
that the Casimir energy can be 
understood from the interaction between surface plasmon resonances on
the two (metallic) mirrors
\cite{VanKampen68}. This holds also in the dissipative case. Adopting
the lossy Drude model 
($\varepsilon=1-\omega^2_p/(\omega^2+\mathrm{i}\gamma\omega)$), one 
can show that Eq.\eqref{eq:sum-modes} reduces to
\begin{equation}
F(L) \approx  \left(\frac{ \alpha \omega_{p} }{ 2\pi }
- \frac{15 \zeta(3) \gamma }{\pi^4} \right)
\frac{\hbar \pi^2 }{240 L^3}
\qquad (\alpha=1.193...),
    \label{eq:plasmons}
\end{equation}
where we have taken the leading order correction in $\gamma$
of Eq.\eqref{eq:sum-modes} and kept in the sum only two modes,
$\omega_{\pm}=\frac{1}{\sqrt{2}} \sqrt{\omega_p^2(1\pm 
e^{-k L})-\gamma^2/2} - \mathrm{i} \gamma/2$,
which are the dissipative counterparts of the coupled surface 
plasmons.

Lifshitz theory also allows to consider materials with engineered
properties. Natural materials have a magnetic permeability which 
actually  can  be set always equal to one in the range of 
frequencies  relevant for the Casimir effect.
Nothing forbids, however, to consider artificial materials (also called 
metamaterials) which show a strong modification of their magnetic 
properties, say, in the visible-light range.
We have recently investigated the simple case of a local 
magneto-dielectric material where both permittivity and permeability
are given by lossy Drude models.\cite{Henkel05a}
More precisely, the permeability is
\begin{equation}
    \mu( \omega ) = 1+\frac{f \omega^2 }{ \omega_0^2 - \omega^2 - 
    \rm{i} \kappa\omega}, \qquad 0 < f < 1.
    \label{eq:eps-mu-metamat}
\end{equation}
Response functions of this kind have been used previously to describe
the response of a metamaterial  to electromagnetic waves.  The 
material contains a regular lattice of
sub-wavelength units (wires and rings) with a size much smaller than
the incident wavelength and filling factor $f$.
With a suitable spatial averaging procedure
(effective medium description)\cite{Ramakrishna05}, one finds the
permeability~(\ref{eq:eps-mu-metamat}).  

The calculation of the Casimir force requires response functions at 
imaginary frequencies. We have used the Kramers-Kronig relation
\begin{equation}
    \mu( {\rm i}Ê\xi ) = 1 + \frac{ 2 }{ \pi }
    \int\limits_{0}^{\infty}\!{\rm d}\omega \, \frac{ \omega \,
    {\rm Im}\, \mu( \omega ) }{ \omega^2 + \xi^2 }
    \label{eq:Kramers-Kronig}
\end{equation}
and focused on the limit of weak absorption where ${\rm Im}\,\mu( 
\omega )$ collapses to a $\delta$-function.
The resulting expression features a ``magnetic plasma frequency''
$\omega_{p} = \omega_{0} \sqrt{f}$.

\begin{figure}
\parbox{6.5cm}{\epsfig{file=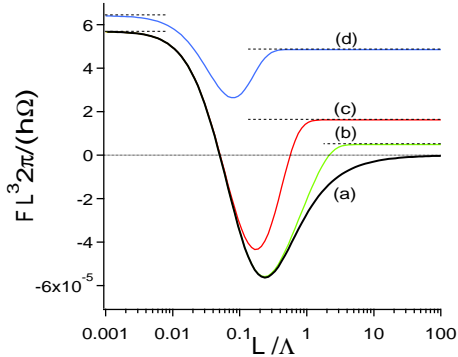,width=6.5cm}}
    \parbox{60mm}{\caption{\small Casimir pressure as a function of 
    distance $L$ between two different metamaterial plates. Positive 
    values correspond to an attractive interaction.
    The force (per unit area) is normalized to $\hbar \Omega / L^3$,
    the distance to $\Lambda \equiv 2\pi c / \Omega$ where $\Omega$ is
    a typical plasma or resonance frequency in
    Eq.(\ref{eq:eps-mu-metamat}).  Plate~1 is purely dielectric,
    plate~2 mainly magnetic.
    The temperature takes the values $k_{B} T = $ (a) $0$, (b)
    $0.03$, (c) $0.1$, (d) $0.3 \,\hbar\Omega$.
    Adapted from Fig.2 of 
    Ref.\citen{Henkel05a} where the parameters can be found.
    }\label{fig:metamat-fig2}}
    \end{figure}

As shown in Fig.\ref{fig:metamat-fig2}, the Casimir interaction
becomes repulsive for a `mixed' pair of mirrors, one mainly
dielectric, the other mainly permeable. This previously discussed
phenomenon\cite{Boyer74} survives in some range of distances
at sufficiently low temperatures even for dispersive materials.
The corresponding parameter window is the wider, the higher the
magnetic plasma frequency.
\vskip -0.5cm

\end{document}